\documentclass[prl,twocolumn,showpacs,showkeys]{revtex4}

\usepackage{amssymb, amsmath, amsfonts, latexsym, verbatim, mathrsfs}

\usepackage{graphicx}

\newtheorem{theorem}{Theorem}

\begin{document}

\title{A general theorem on local parts of hidden variable models}
\thanks{One of us, R.R., acknowledges the financial support by the R. Parisi Foundation }

\author{GianCarlo Ghirardi}
\email{ghirardi@ictp.it}
\affiliation{Department of Physics, University of Trieste, The Abdus Salam ICTP, Trieste, Italy}
\author{Raffaele Romano}
\email{rromano@ts.infn.it}
\affiliation{Department of Physics, University of Trieste, Fondazione Parisi, Rome, Italy}


\begin {abstract}
\noindent We extend to any maximally entangled state of a bipartite system whose constituents are arbitrarily (but finite) dimensional the result, recently derived for two-dimensional constituents, that  hidden variable theories cannot have local parts, i.e., that their local averages cannot differ from the quantum mechanical ones.
\end{abstract}



\maketitle


{\it Introduction ---} Quantum mechanics offers a very accurate description of the microscopic word, as shown by the astonishing agreement between its predictions and experimental data. Nonetheless, it suffers from some conceptual problems,  concerning mainly the extension of the formalism to macroscopic objects and the interaction between micro- and macro-systems. Moreover, it offers a picture of reality which is rather counterintuitive: the statistical character of the formalism and the uncertainty relations are prominent examples of this fact. For these reasons, it has been repeatedly investigated whether one could devise a more complete description of  reality making less problematic all (or just some) of its non-classical aspects. The hope is that the quantum features might be accounted in terms of our ignorance of some hidden parameters. This idea, introduced in the famous paper on EPR correlations \cite{Einstein} arguing that quantum mechanics is an incomplete theory, has lead to the development of the so-called hidden variables program, that is, the elaboration of theories which are predictively equivalent to quantum mechanics but differ from it both for formal as well as for conceptual aspects.  A prominent example of hidden variables theory is given by  Bohmian mechanics \cite{Bohm}, in which, differently from quantum mechanics, the particles possess at every time a well defined position even though it is not accessible. Noticeably, the development of hidden variables theories has played an important role in understanding much better what are presently considered the fundamental traits of quantum mechanics, i.e., entanglement, non-locality and contextuality \cite{Bell}.

The most common hidden variables theories which have been considered  assume  that the state of a system is not fully described by the quantum state vector $\psi$, but additional variables $\lambda$, besides $\psi$, are needed. These variables are not accessible, and their distribution $\rho (\lambda)$ is responsible for the statistical spreads implied by quantum mechanics. We will follow such a line of thought for the case of a bipartite scenario, that is, when the system is shared between two parties, Alice and Bob, which can perform local measurements on the subsystem they own. We assume that the Hilbert spaces of the subsystems have the same dimensionality $N$. Generic local observables pertaining to Alice and Bob are denoted by $A$ and $B$, and, as well known, in quantum mechanics they are associated to  hermitian operators $\hat{A}$ and $\hat{B}$, respectively. We will identify the specific local observables we will deal with by specifying (see below for their precise characterization)  a set of $N^{2}$ real parameters $a$ for $A$ and $b$ for $B$. Accordingly, we will indicate the observables we will be dealing with as  $A(a)$ and $B(b)$, (and the corresponding quantum operators by $\hat{A}(a)$ and $\hat{B}(b)$).  As usual, we will assume that, once the state vector of the composite system $\psi$ and ``the settings" $a$  and $b$ are specified, the considered observables take precise values $A_{\psi}(a,b,\lambda)$ and $B_{\psi}(a,b,\lambda)$ which  belong to the spectra of $\hat{A}(a)$ and $\hat{B}(b)$, respectively. If $\{A(a), B(b) \}$ is not a complete set of commuting observables and $N > 2$, these values in general depend also on additional parameters which specify the physical context \footnote{To be more precise, to avoid the ambiguity arising from contextuality, we should use the notation $A_{\psi}(a,a^{\prime},\ldots,b,b^{\prime},\ldots,\lambda)$ and $B_{\psi}(a,a^{\prime},\ldots,b,b^{\prime},\ldots,\lambda)$, where $a, a^{\prime}, \ldots$ and $b, b^{\prime}, \ldots$ specify sets of commuting observables for the two subsystems which also commute with $\hat{A}(a)$ and $\hat{B}(b)$ and, with them, constitute a complete set. However, for sake of simplicity, we prefer a notation which makes explicit nonlocality and not contextuality.}. The joint dependence on $a$ and $b$ is an unavoidable signature of non-locality; consistency with quantum statistics requires:

\begin{eqnarray}\label{hvav}
    \langle A(a) \rangle_{\psi} &=& \int A_{\psi} (a,b,\lambda) \rho (\lambda) d \lambda, \nonumber \\
    \langle B(b) \rangle_{\psi} &=& \int B_{\psi} (a,b,\lambda) \rho (\lambda) d \lambda, \\
    \langle A(a) B(b) \rangle_{\psi} &=& \int A_{\psi} (a,b,\lambda) B_{\psi} (a,b,\lambda) \rho (\lambda) d \lambda, \nonumber
\end{eqnarray}
where $\langle A(a) B(b) \rangle_{\psi} = \langle \psi \vert \hat{A}(a) \otimes \hat{B}(b) \vert \psi \rangle$, and the local averages $\langle A(a) \rangle_{\psi}$, $\langle B(b) \rangle_{\psi}$ are defined by choosing $\hat{B}(b) = \hat{I}$, $\hat{A}(a) = \hat{I}$  in the previous expression, respectively. In (\ref{hvav}) we have tacitly assumed that $\lambda$ are continuous variables; the conclusions we will draw do not depend on this assumption.

It is a physically relevant fact that non-locality can only be detected by considering joint measurements performed by Alice and Bob, in particular, by comparing them. In fact, as implied by the first two of the above equations,  when a local measurement is taken into account, the average over $\lambda$ completely washes out the dependence on the setting of the other party. It is appropriate to recall that this  is a necessary condition to guarantee the peaceful coexistence of the theory with the causality principle. Nevertheless, one could impose this requirement at a different level, for instance, by splitting the hidden variables in two families, $\lambda = (\mu, \tau)$, with the integration over $\mu$ already washing out non-locality and contextuality of local observables:
\begin{eqnarray}\label{cnlhv}
     \int A_{\psi} (a,b,\mu,\tau) \rho (\mu \vert \tau) d \mu,&=&  f_{\psi}(a,\tau), \nonumber \\
     \int B_{\psi} (a,b,\mu,\tau) \rho (\mu \vert \tau) d \mu &=& g_{\psi}(b,\tau),
\end{eqnarray}
where $\rho (\lambda) = \rho (\mu \vert \tau) \rho (\tau)$. In such a case the quantum averages over $\psi$ of local observables could be associated with a local classical distribution, since
\begin{eqnarray}\label{cnlhv2}
  \langle A(a) \rangle_{\psi} &=& \int f_{\psi}(a,\tau) \rho (\tau) d \tau, \nonumber \\
  \langle B(b) \rangle_{\psi} &=& \int g_{\psi}(b,\tau) \rho (\tau) d \tau.
\end{eqnarray}
These ideas have been introduced by Leggett \cite{Leggett}, in the attempt of interpreting, at the intermediate level of the hidden variables $\mu$, the singlet state of two polarized photons as associated to an ensemble of pairs of photons emitted by the source with definite polarizations. Since this author was investigating the $N = 2$ case he had not to face the problem of the possible contextual nature of single particle observables. However, it perfectly fits his line of thought the assumption, which we have made explicit in (\ref{cnlhv}), that the average over $\mu$ washes out also the possible contextuality of the quantities $A_{\psi} (a,b,\mu,\tau)$ and $B_{\psi} (a,b,\mu,\tau)$. 

Obviously, when $\psi$ is the singlet state and $A$, $B$ are polarization observables,  requirement (2) is physically meaningful only if, differently from the quantum case, $f_{\psi}(a,\tau) \ne 0$ and $g_{\psi}(b,\tau) \ne 0$. The explicit choice made by Leggett for the functions $f_{\psi}(a,\tau)$ and $g_{\psi}(a,\tau)$ leads to a disagreement with quantum predictions and with experimental data, as noted by Leggett himself, and shown in \cite{Groblacher}. Nonetheless, general models satisfying (\ref{cnlhv}), called {\it crypto-nonlocal}, have been the subject of further investigations, to check whether it is possible to attach local properties to the maximally entangled states of two qubits, although only at the intermediate level of the hidden variables.

Actually, in an interesting series of papers \cite{Gisin, Colbeck, Parrott, Colbeck2} it has been proved that consistency with quantum mechanics necessarily implies $f_{\psi}(a,\tau) = g_{\psi}(b,\tau) = 0$ when $\psi$ is a maximally entangled state of a system of two qubits, so that $A({\bf a})=\sigma^{(A)}\cdot{\bf a}$, and $B({\bf b}) = \sigma^{(B)} \cdot {\bf b}$ have spectrum $\{-1, 1\}$. This being the case, one cannot take advantage of the crypto-nonlocal approach to attribute some individual properties to the constituents.

In this letter, we generalize this result, and prove that,  for an arbitrary maximally entangled state of a pair of two $N$-level systems and for completely generic observables $A$ and $B$, one must always have: $f_{\psi}(a,\tau) = \langle A(a) \rangle_{\psi}$ and $g_{\psi}(b,\tau) = \langle B(b) \rangle_{\psi}$. Already  for the case $N = 2$, our derivation is more general than those which can be found in the literature, as we will point out below.

{\it Preliminary results} --- In the following, $\psi$ denotes a maximally entangled state of a system of two $N$-level systems ($N \geqslant 2$) whose Schmidt decomposition, in Dirac notation, reads
\begin{equation}\label{schmidt}
    \vert \psi \rangle = \frac{1}{\sqrt{N}} \sum_{j = 1}^N \vert v_j \rangle \otimes \vert w_j \rangle,
\end{equation}
where $\{\vert v_j \rangle; j\}$ and $\{\vert w_j \rangle; j\}$ are orthonormal bases for ${\mathcal H}_A$ and ${\mathcal H}_B$, the Hilbert spaces pertaining to Alice and Bob, respectively. The Hilbert space of the composite system is  ${\mathcal H} = {\mathcal H}_A \otimes {\mathcal H}_B$.

Given a maximally entangled state $\vert\psi\rangle$ of the composite system, for every local operator $\hat{O} = \sum_{ij} o_{ij} \, \vert v_i \rangle \langle v_j \vert$ there is an associated local operator $\hat{O}^T = \sum_{ij} o_{ji} \, \vert w_i \rangle \langle w_j \vert$, of the other subsystem, such that
\begin{equation}\label{propme}
    \hat{O} \otimes \hat{I} \, \vert \psi \rangle = \hat{I} \otimes \hat{O}^T \vert \psi \rangle.
\end{equation}
Having in mind this property, we introduce  $\psi$-dependent bases for the sets of the Hermitian operators pertaining to Alice and Bob according to: 
\begin{eqnarray}\label{basis}
    \hat{F}_{ii} &=& \vert v_i \rangle \langle v_i \vert, \nonumber \\
    \hat{F}^{+}_{ij} &=& \frac{1}{\sqrt{2}} \, (\vert v_i \rangle \langle v_j \vert + \vert v_j \rangle \langle v_i \vert), \\
    \hat{F}^{-}_{ij} &=& \frac{i}{\sqrt{2}} \, (\vert v_i \rangle \langle v_j \vert - \vert v_j \rangle \langle v_i \vert), \nonumber
\end{eqnarray}
with $i < j = 2, \ldots, N$. We rearrange these Hermitian operators in a vector $\hat{F} = (\hat{F}_k; k = 1, \ldots, N^2)$, and write $\hat{A}(a) = a \cdot \hat{F}$, where $\cdot$ denotes, as usual, the inner product in the $N^{2}$ dimensional space to which $a\equiv (a_{ii},a_{ij}^{+},a_{ij}^{-})$  belongs. Notice that ${\rm Tr} (F_k F_l) = \delta_{kl}$ by construction. The corresponding basis on the Bob side is defined as $\hat{G} = \hat{F}^T$, and the generic  observable of Bob takes the form $\hat{B}(b) = b \cdot \hat{G}$. Note that Hermiticity requires that both $a$ and $b$ are real, $N^{2}$-dimensional vectors. These are the parameters which we have mentioned above as characterizing uniquely the observables we will deal with. With these assumptions, taking advantage of the property (\ref{propme}), the joint quantum averages assume the particularly simple form
\begin{equation}\label{avjo}
    \langle A(a) B(b) \rangle_{\psi} = \frac{1}{N} \, a \cdot b,
\end{equation}
and moreover
\begin{equation}\label{avsq}
     \langle A(a)^2 \rangle_{\psi} = \frac{1}{N} \, \Vert a \Vert^2, \quad \langle B(b)^2 \rangle_{\psi} = \frac{1}{N} \, \Vert b \Vert^2.
\end{equation}
For further reference we observe that, if $A(a)$ and $B(b)$ have vanishing averages over $\psi$, the Pearson correlation coefficient $r(A,B)$ turns out to have the following expression:
\begin{equation}\label{corr}
    r(A,B) = \frac{\langle A(a) B(b) \rangle_{\psi}}{\sqrt{\langle A^2(a) \rangle_{\psi} \langle B^2(b) \rangle_{\psi}}} = \frac{a}{\Vert a \Vert} \cdot \frac{b}{\Vert b \Vert}.
\end{equation}
Therefore for $a = b$ ($a = - b$) the observables $A(a)$ and $B(b)$ are completely correlated (anti-correlated).

We  consider now an arbitrary  observable $A(a)$ of Alice and its associated operator $\hat{A}(a)$. By means of the Cartan decomposition of the Lie algebra of Hermitian operators, we can write
\begin{equation}\label{gena}
    \hat{A}(a) = \alpha_0 \hat{I} + \sum_{j = 1}^{N - 1} \alpha_j \hat{A}(a_j),
\end{equation}
where $\alpha_j$ are real coefficients for all $j = 0, \ldots N - 1$, and $\{ \hat{A}(a_j); j \}$ is a set of commuting Hermitian operators, with the following property. When $N > 2$, they have spectrum $\Omega_N = \{ -1, 0, 1 \}$, with the null eigenspace (of dimension $N - 2$) as the only degenerate one; if $N = 2$, the spectrum is $\Omega_2 = \{ -1, 1 \}$ without degeneration. The set $\{ \hat{A}(a_j);j \}$ depends on $a$, and it is not a basis for the space of Hermitian operators, which requires $N^2$ operators. Since ${\rm Tr} \, \hat{A}(a_j) = 0$ for all $j$, it follows that
\begin{equation}\label{ava}
    \langle A(a) \rangle_{\psi} = \frac{1}{N} \, {\rm Tr} \, \hat{A}(a) = \alpha_0.
\end{equation}
The decomposition (\ref{gena}) is not uniquely defined. By using the spectral theorem, it is possible to prove that, if $\hat{A}(a)$ in non-degenerate, operators appearing in different decompositions  commute. The situation is different when $\hat{A}(a)$ is degenerate: in such a case it always admits different decompositions involving non-commuting operators.

Note that any hidden variables theory predictively equivalent to quantum mechanics must assign a precise value to the observables $A(a)$ and $A(a_j)$ for all $j$. Moreover, the values of the observables  must satisfy the analogue of the functional relation (\ref{gena}):
\begin{equation}\label{ass}
    A_{\psi} (a,b,\lambda) = \alpha_0 + \sum_{j = 1}^{N - 1} \alpha_j A_{\psi} (a_j,b,\lambda),
\end{equation}
where $A_{\psi}(a_j,b,\lambda)$ are associated to the observables $A(a_j)$ corresponding to $\hat{A}(a_j)$. Notice that, if $\hat{A}(a)$ is degenerate, for $N > 2$ the theory must be contextual: it is impossible to unambiguously impose (\ref{ass}) for every possible decomposition of $\hat{A}(a)$. However, when the physical context is specified (corresponding to a complete set of commuting observables to which $\hat{A}(a)$ belongs), no ambiguity arises, since the decomposition (\ref{gena}) in this case is fixed: the observable $\hat{A}(a)$ as well as the $\hat{A}(a_{j})$ are functions of the complete set defining the context and, as such, they have precise values once $\lambda$ is specified.

In a crypto-nonlocal theory $\lambda = (\mu, \tau)$, and, in accordance with (\ref{cnlhv}), at the intermediate level we find
\begin{equation}\label{intf}
    f_{\psi} (a, \tau) = \langle A(a) \rangle_{\psi} + \sum_{j = 1}^{N - 1} \alpha_j f_{\psi} (a_j, \tau),
\end{equation}
where we have expressed $\alpha_0$ through (\ref{ava}). A completely analogous analysis can be performed for  Bob's observables, so that we can conclude that
\begin{equation}\label{intg}
    g_{\psi} (b, \tau) = \langle B(b) \rangle_{\psi} + \sum_{j = 1}^{N - 1} \beta_j g_{\psi} (b_j, \tau),
\end{equation}
where $\beta_j$ and $b_j$ are the analogues of $\alpha_j$ and $a_j$ respectively.

We now limit our attention to those observables $A(a)$ and $B(b)$ of Alice and of Bob, respectively,  such that $\langle A(a) \rangle_{\psi} = \langle B(b) \rangle_{\psi} = 0$. As we are going to show one has: $f_{\psi} (- a, \tau) = - f_{\psi} (a, \tau)$, and the same for $g$. Since $a$ is generic, we conclude that both $f_{\psi}$ and $g_{\psi}$ are skew-symmetric in their arguments.

{\it Proof -} The observables $A(a)$ and $B(a)$ are perfectly correlated and they have the same spectrum, then $A_{\psi} (a,a,\mu,\tau) = B_{\psi} (a,a,\mu,\tau)$. Averaging over $\mu$ leads to $f(a, \tau) = g(a, \tau)$. Repeating the argument for $A(a)$ and $B(-a)$, which are perfectly anti-correlated, we find that $f_{\psi}(a, \tau) = -g_{\psi}(-a, \tau)$. The proof is concluded by suitably combining the two relations.

{\it Main result ---} We further restrict our attention to operators with the aforementioned spectrum $\Omega_N$, $N \geqslant 2$, and consider Alice's observables for simplicity. We find convenient to write ${\mathcal H}_A = {\mathcal K} \oplus {\mathcal L}$, where ${\mathcal K}$ is the kernel of $\hat{A}(a)$, and ${\mathcal L}$ its orthogonal space, with ${\rm dim} \, {\mathcal L} = 2$. Of course, if $N = 2$, ${\mathcal H} = {\mathcal L}$. We observe that
\begin{equation}\label{pama}
    \hat{A} (a)\vert_{\mathcal L} = \tilde{a} \cdot \hat{\sigma}, \quad \hat{A} (a)\vert_{\mathcal K} = \hat{0},
\end{equation}
where $\hat{\sigma}$ is the vector of Pauli matrices acting on ${\mathcal L}$, and $\hat{0}$ is the null operator acting on ${\mathcal K}$. The vector $\tilde{a}$ is a $3$-dimensional real vector of unit length. Notice that this vector, as well as the subspaces ${\mathcal K}$ and ${\mathcal L}$, depend on $a$ and are univocally determined by it, when a specific context is assigned. The explicit form of this dependence is not relevant for what follows.


For future reference, it is useful  to build a curve in the $N^2$-dimensional real space, connecting $a$ to $-a$. With this  in mind, we define unitary operators $\hat{V}_{\theta} \in SU(N)$, acting on ${\mathcal H}$, such that
\begin{equation}\label{unit}
    \hat{V}_{\theta} \vert_{\mathcal L} = \hat{U}_{\theta}, \quad \hat{V}_{\theta} \vert_{\mathcal K} = \hat{I,}
\end{equation}
where $\hat{U}_{\theta} \in SU(2)$ acts on ${\mathcal L}$, and $\theta \in [0, \pi]$. We assume that $\hat{U}_{\theta} = e^{i \frac{\theta}{2} \tilde{c} \cdot \hat{\sigma}}$, where $\tilde{c}$ is a $3$-dimensional real vector such that $\tilde{c} \cdot \tilde{a} = 0$. It turns out that
\begin{equation}\label{transf}
    \hat{U}_{\theta} \, \tilde{a} \cdot \hat{\sigma} \, \hat{U}_{\theta}^{\dagger} = \tilde{a}(\theta) \cdot \hat{\sigma},
\end{equation}
where $\tilde{a}(\theta) = R (\theta) \, \tilde{a}$, and $R(\theta)$ is a $SO(2)$ rotation. We define the aforementioned curve as the one-parameter family of vectors $\gamma = \{ a(\theta); 0\leqslant \theta \leqslant \pi \}$, where $a(\theta)$ is defined by
\begin{equation}\label{gentra}
    \hat{V}_{\theta} \hat{A}(a) \hat{V}_{\theta}^{\dagger} = \hat{A}(a(\theta)).
\end{equation}
By construction, we have $a(0) = a$ and $a(\pi) = -a$. Since $V_{\theta}$ is a unitary transformation, it follows that $\Vert a(\theta) \Vert = \Vert a \Vert$ by (\ref{avsq}), and $\hat{A} (a(\theta))$ has spectrum $\Omega_N$, and then null trace, for every $\theta \in [0, \pi]$. Finally, we notice that the decomposition ${\mathcal K} \oplus {\mathcal L}$ is independent of $\theta$, therefore the action of $\hat{V}_{\theta}$ is non-trivial only in ${\mathcal L}$ for all $\theta$. This property implies that $\gamma$ is a planar curve.
We are ready to prove the fundamental theorem.

\begin{theorem}\label{theo1}
 Consider the maximally entangled state $\psi \in {\mathcal H} = {\mathcal H}_A \otimes {\mathcal H}_B$, with ${\rm dim} \, {\mathcal H}_A = {\rm dim} \, {\mathcal H}_B = N \geqslant 2$, and the observable $A (a)$, associated to the Hermitian operator $\hat{A} (a)$ with spectrum $\Omega_N$. Then necessarily $f_{\psi}(a, \tau) = 0$.
\end{theorem}

{\it Proof -} For a fixed natural number $n$, we define $\theta_j = \frac{j}{n} \, \pi$, with $j = 0, \ldots, n$, such that $\theta_0 = 0$, $\theta_n = \pi$. For simplicity, we write $a_j = a(\theta_j)$, where $a(\theta) \in \gamma$. It turns out that
\begin{equation}\label{spacing}
    a_{j + 1} \cdot a_j = \Vert a \Vert^2 \cos{\frac{\pi}{n}}, \quad j = 0, \ldots, n - 1.
\end{equation}
We consider the setups in which the  measurement by Alice is determined by $a_j$ and the one of Bob  by $a_{j + 1}$, for $j = 0, \ldots, n - 1$. Since both $\hat{A}(a_j)$ and $\hat{B}(a_{j + 1})$ have spectrum $\Omega_N$, we can write
\begin{eqnarray}\label{proprel}
    &&\vert A_{\psi}(a_j,a_{j + 1},\lambda) - B_{\psi}(a_j,a_{j + 1},\lambda) \vert \\
    && \quad \leqslant \Bigl(A_{\psi}(a_j,a_{j + 1},\lambda) - B_{\psi}(a_j,a_{j + 1},\lambda)\Bigr)^2 \nonumber
\end{eqnarray}
for all $j$. If we multiply (\ref{proprel}) by $\rho(\lambda)$ and integrate it over $\lambda$, by considering elementary properties of integrals, and making use of (\ref{avjo}), (\ref{avsq}), (\ref{spacing}), and the crypto-nonlocality conditions (\ref{cnlhv}), we obtain
\begin{equation}\label{propel2}
    \int \vert f_{\psi}(a_j, \tau) - g_{\psi}(a_{j + 1}, \tau) \vert \rho (\tau) d \tau \leqslant \frac{4 \Vert a \Vert^2}{N} \sin^2{\frac{\pi}{2n}}
\end{equation}
for all $j$. At the l.h.s., we use the fact that $g_{\psi} (a_{j + 1}, \tau) = f_{\psi}(a_{j + 1}, \tau)$, and then we sum these expressions for $j = 0, \ldots, n - 1$, and use the triangle inequality to obtain:
\begin{equation}\label{propel3}
    \int \vert f_{\psi}(a_0, \tau) - f_{\psi}(a_{n}, \tau) \vert \rho (\tau) d \tau \leqslant \frac{4 n \Vert a \Vert^2}{N} \sin^2{\frac{\pi}{2n}}.
\end{equation}
Now, since $a_0 = a$, $a_n = -a$, and $f_{\psi}$ is skew-symmetric for $a \rightarrow -a$, we conclude that
\begin{equation}\label{propel4}
    \int \vert f_{\psi}(a, \tau) \vert \rho (\tau) d \tau \leqslant \frac{2 n \Vert a \Vert^2}{N} \sin^2{\frac{\pi}{2n}}.
\end{equation}
Note that, for $n \rightarrow + \infty$, the expression at the r.h.s. of (23) vanishes. Therefore $f_{\psi}(a, \tau) = 0$ almost everywhere. QED

This result applies as well to Bob observables, and it implies that $f_{\psi} (a_j, \tau) = g_{\psi} (b_j, \tau) = 0$ in (\ref{intf}) and (\ref{intg}). Therefore, for arbitrary $A(a)$ and $B(b)$ we conclude that $f_{\psi}(a,\tau) = \langle A(a) \rangle_{\psi}$ and $g_{\psi}(b,\tau) = \langle B(b) \rangle_{\psi}$.

We notice that the curve which connects $a$ to $-a$ is not unique, since there are infinitely many vectors $\tilde{c}$ orthogonal to $\tilde{a}$. Moreover, it is not really necessary to use $V_{\theta}$ defined in (\ref{unit}), a more general family of unitary operators works as well as long as the corresponding curve has $a$ and $-a$ as extremal points. In such a case, this curve is not planar, and then the partition defined by (\ref{spacing}) has to be replaced by one with
\begin{equation}\label{spacing2}
    a_{j + 1} \cdot a_j = \Vert a \Vert^2 \cos{\frac{\Theta}{n}}, \quad j = 0, \ldots, n - 1,
\end{equation}
where $\Theta > \pi$. The argument and conclusions of Theorem \ref{theo1} hold also in this case.

{\it Conclusions ---} In this work we have considered an arbitrary crypto-nonlocal hidden variable theory describing a bipartite system in a maximally entangled state, with arbitrary dimension for the Hilbert spaces of the two subsystems. This constitutes an interesting generalization of the theorem derived in \cite{Gisin,Colbeck} for $N = 2$. We have proven that the averages produced by this theory at the intermediate level of the hidden variables cannot differ from the quantum mechanical averages. In the special case of qubits, our derivation differs from the former derivations, and is more general than them, in that: (i) it holds for arbitrary local operators; (ii) no assumptions are made on the nature of the hidden variables or they distribution, which could depend on the  local settings.

The properties of quantum systems are often highly dependent on the state space dimension. We have proved that this is not the case for the intermediate averages of a crypto-nonlocal hidden variable theory, as long as this dimension is finite. In our opinion, this fact does not exhaust the interest in these theories, as implied e.g. by the fact that one can build a theory of this type for which the $\mu$-averages of the local observables actually agree with the quantum expectation values while the same average of the correlations can overcome the upper bound of $2\sqrt{2}$ which, within quantum mechanics, corresponds to the maximal possible violation of locality  \cite{Ghirardi}. Accordingly, their consideration could possibly push forward our understanding of the general features and the conceptual structure, from the point of view of the locality issue, of hidden variables theories equivalent to quantum mechanics.


\end{document}